\documentclass[oldversion]{aa}
\usepackage{graphicx}
\usepackage{txfonts}
\begin{document}

\titlerunning{Integral field spectroscopy in the near infrared}
\authorrunning{Vanzi et al.}

\title{Integral field spectroscopy in the near infrared of NGC 3125-A and SBS~0335-052\thanks{Based on observations obtained at the ESO-VLT under program 381.B-0132.}}

\author{L. Vanzi \inst {1},
          G. Cresci     \inst{2},
          M. Sauvage \inst{3},
          \and 
          R. Thompson \inst{4}.
          }

\offprints{L. Vanzi}
 \institute{Department of Electrical Engineering and Center of Astro Engineering, Pontificia Universidad Catolica de Chile, \\
 Av. Vicu\~{n}a Mackenna 4860, Santiago, Chile\
 \email{lvanzi@ing.puc.cl}
 \and
 INAF - Osservatorio Astrofisico di Arcetri, Largo E. Fermi 5, 50125 Firenze, Italy
 \and
  Laboratoire AIM, CEA, UniversitŽ Paris Diderot, IRFU/Service d'Astrophysique, B‰t. 709, 91191, Gif-sur-Yvette, France\
  \and
   Steward Observatory, University of Arizona, Tucson, AZ 85721, USA }

   \date{Received 2 June 2011 / accepted 26 July 2011 }

   \abstract{We present integral field spectroscopy in the near infrared of the nearby dwarf starburst galaxies NGC 3125-A and of the low metallicity dwarf galaxy SBS~0335-052. The use of adaptive optics in the observations produces sub-arcsecond angular resolution.
 We pinpoint the star forming cores of both galaxies, identify relevant ISM components such as dust, photo ionized gas, shock excited gas and molecular gas. We relate these components to the large scale star formation process of the galaxies. In particular we find the emission of the near infrared lines of $H_2$ and especially [FeII] does not  coincide with the HII region in NGC~3125. We have the first clear detection of [FeII] in SBS~0335-052.   
   
}

   \maketitle

\section{Introduction}

In order to understand the processes that affect star formation on galactic scales, galaxy assembly, and evolution, we began a survey campaign to observe a number of nearby dwarf galaxies using near infrared integral field spectroscopy.
In this way we have access to a number of unique features that characterize the star formation environment.
 As part of this investigation we have already published results on three relevant objects: II~Zw~40 (Vanzi et al. 2008), He~2-10 and NGC~5253 (Cresci et al. 2010).
All these galaxies are nearby irregular blue dwarfs undergoing powerful episodes of star formation, as such they reveal the details star formation on a large scale.

Here we report new observations of two other examples of this class of 
object:
 NGC~3125 and SBS~0335-052. We used AO correction to reach sub-arcsec angular resolution in our observations and  in order to pinpoint the star forming cores of the galaxies.

Star formation in NGC~3125 (also known as Tol~3) occurs in a number of compact regions of which the most prominent ones were first indicated as 2 and 3 by Kunth et al. (1988) later renamed A and B by Vacca \& Conti (1992) and Stevens et al. (2002). Those regions are separated by about 10 arcsec in a direction defining a PA=124$^{\circ}$. Both regions are 
further separated into two clusters respectively in HST images (see Hadfield \& Crowther 2006). Clusters A1 and A2 are separated by about 0.5 arcsec, with cluster A1 dominating the emission in the HST/FOC F220W filter; clusters B1 and B2 are separated by about 0.2 arcsec and have similar luminosities in the mentioned band.
NGC~3125 has characteristics similar to the three galaxies previously observed by us: relatively low mass, high star formation rate per unit mass, presence of young massive clusters, WR features and irregular morphology. NGC~3125, however, differs from them in the radio continuum.
Stevens et al. (2002) observed  NGC 3125 in the radio at 6 and 3 cm and resolved the two regions A and B. Both regions show a negative spectral index indicating the predominance of non thermal radiation at those wavelengths.
II~Zw~40, He~2-10, and NGC~5253 instead are characterized by mostly thermal (bremstrahlung) radio spectra (Beck et al. 2002, Jonhson \& Kobulnicky 2003, Turner et al. 1998). Spectra of NGC~3125 in the near infrared were observed by Vanzi et al. (2002), they found bright emission lines of H, He, and H$_2$ in 3125-A, much fainter lines in 3125-B. 
The metallicity is slightly sub-solar, Engelbracht et al. (2008) report 12+Log(O/H) = 8.34, about a factor 3 lower than in the sun. At a distance of 11.5 Mpc, 1 arcsec is equivalent to 55.7 pc. Because of our limited field of view we concentrated our observations on region A.

SBS~0335-052 was subject of several studies and observations in the past. Its interest resides mostly in the low metal content of the interstellar gas that was measured by Melnick et al. (1992) and Izotov et al. (1997) and found to be about 1/40 of the solar value. Despite initial thoughts, the idea that this galaxy might be a primeval object has been doubted (\"{O}stlin \& Kunth 2001) and further weakened  by the finding of a population of evolved stars in I~Zw~18 a galaxy closer to us but characterized by even lower metallicity (\"{O}stlin \& Mouhcine 2005, Aloisi et al. 2007). 
The stellar population in SBS~0335-052 cannot be resolved due to its large distance.%

HST observations by Thuan et al. (1997) indicated that in this object star formation is occuring in a number of super star clusters, spanning a range of ages, the area covered by these clusters is about 2" in length. 
The distance to SBS-0335-052 of 54.3 Mpc (Thuan et al. 1997) is significantly larger than for NGC-3125, however the region of star forming activity in SBS-0335-052 has a linear size very similar to the distance between region A and B in NGC~3125, i.e. 526 pc.
 The galaxy was observed in the near infrared by Vanzi et al. (2000) and more recently by Thompson et al. (2006, 2009) with NICMOS on board HST. The optical clusters are also detected in the near IR and
 Thompson et al. (2009) find additional regions of very recent star formation not detected previously. 
Observations in the radio were obtained by Hunt et al. (2004) finding a composite thermal plus nonthermal slope.

This paper contains six main Sections, in Sect. 2 we present our new observations and data reduction, Sect. 3 is dedicated to the results on NGC~3125, Sect. 4 to the results on SBS~0335-052. Sect. 3 and 4 are organized in five main parts: near IR continuum, ionized hydrogen, dust and extinction, molecular hydrogen, and ionized iron. In Sect. 5 we discuss the results of our work. Sect. 6 is the summary.


\section{Observations}

The observations were obtained with SINFONI at the ESO-VLT. SINFONI is a 3D near-IR spectrograph with AO correction (Eisenhauer et al. 2003).

For NGC~3125 we used the 0.250$\times$0.125 arcsec/pixel scale which provides a field of view of 8 arcsec. For SBS~0335-052 we used the 0.100$\times$0.050 arcsec/pixel with a field of view of 3 arcsec.
The observations were performed with AO correction and laser guide star, this allowed to reach an angular resolution of approximately  0.25  arcsec, measured deconvolving the FWHM of clusters with their angular size in HST images. We obtained spectra with the H and K grism, with an integration time of 20 minutes each and resolution of 2000 and 3000 respectively. 

The data reduction was performed with the ESO pipeline and followed the standard steps of flatfielding,  correction for dead/hot pixels, background subtraction and telluric correction. Flux calibration were carried out using A- and B-type stars, after removing the stellar features. Residuals from the OH line emission were minimized with the methods outlined in Davies (2007). 

We compared both aperture photometry and emission line fluxes for the sources present in our fields that have previous observations in the literature and found good agreement in most cases (see Sect. 3.1 and 4.1), this makes us confident of the good level of  reliability of our measures.

Integral field spectroscopy produces spectra for each pixel in our FOV thus providing spectral images at all observed wavelengths.
Our analysis concentrated on a few features visible in the spectra, mainly the emission lines of atomic hydrogen, the emission lines of molecular hydrogen, the forbidden emission lines of ionized iron [FeII], and the nearby continuum. We extracted spectral images centered on these features. This was done by fitting a function to the continuum-subtracted spectral profile at each spatial position in the datacube.
The function fitted was a convolution of a Gaussian with a spectrally unresolved emission line profile of a suitable sky line. The parameters of the Gaussian were adjusted until the convolved profile best matched the data, see Cresci et al. (2009) for further details.

\section{Analysis of NGC~3125}

\begin{figure*}
\centering
\includegraphics[angle=0, width=19cm]{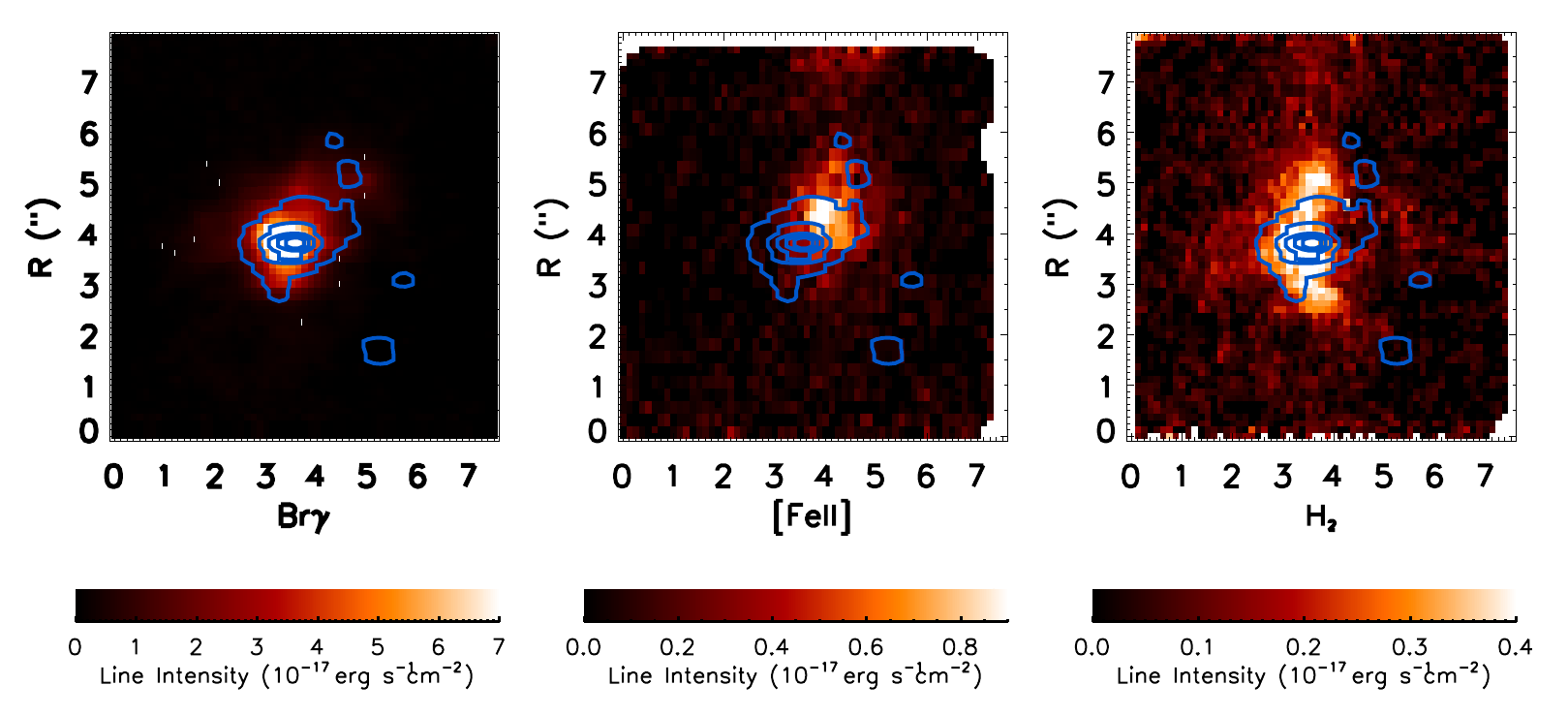}
\caption{Br$\gamma$, [FeII] and $H_2$ images of NGC~3125, the contours of the continuum are over plotted on each image. The continuum plotted corresponds to band H for [FeII] and to band K for Br$\gamma$ and $H_2$.}
\label{flux_3125}
\end{figure*}

In Fig. \ref{flux_3125}  we show the images in Br$\gamma$ 2.167 $\mu$m, [FeII] 1.643 $\mu$m, and$H_2$ 2.121 $\mu$m rest wavelengths of NGC~3125-A with the contours of the continuum  overplotted for reference. We will base  our main analysis on these three images. 

\subsection{Near Infrared continuum}
\begin{figure*}
\centering
\includegraphics[angle=0, width=15cm]{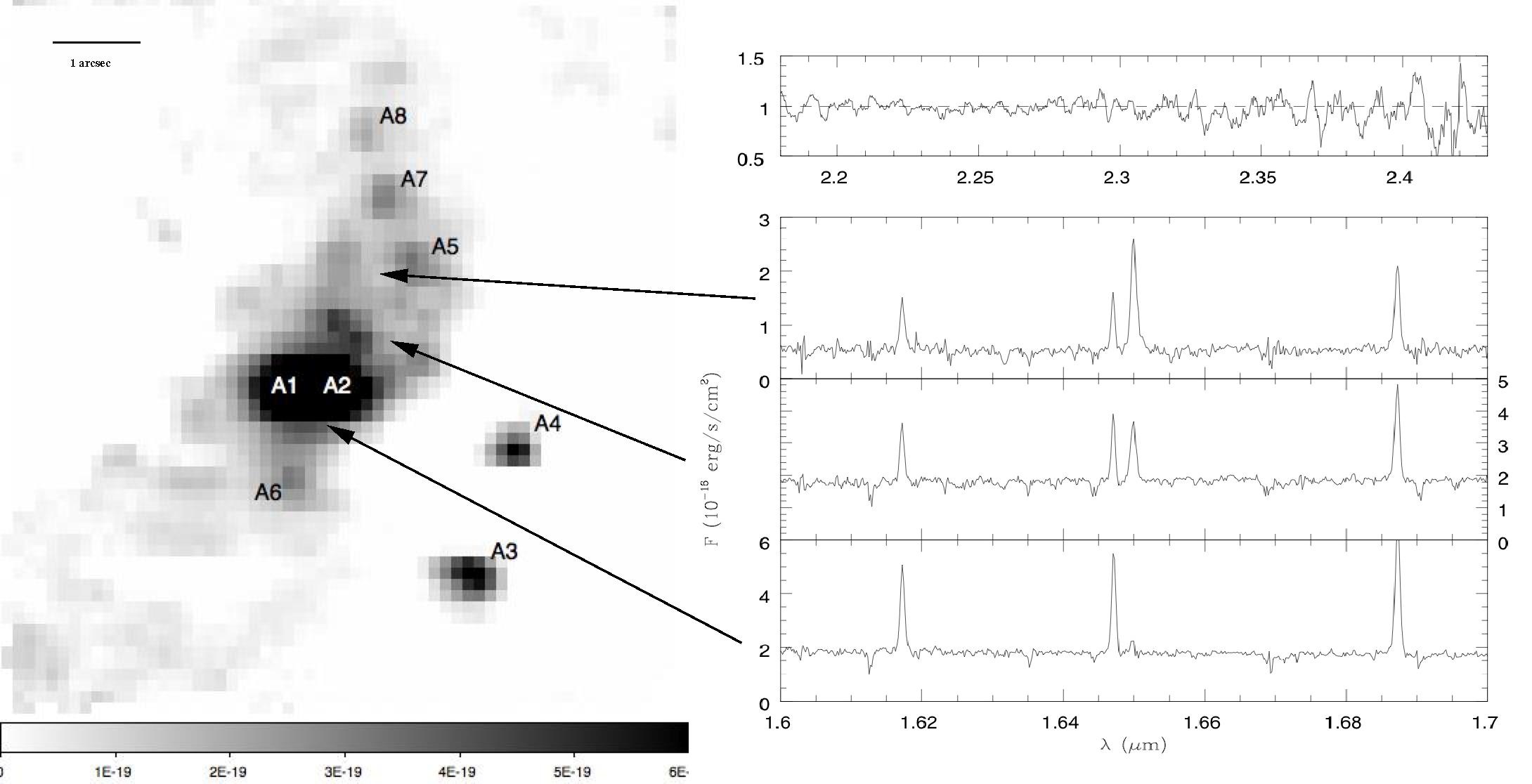}
\caption{In the left panel NGC~3125 in the K continuum, the compact sources detected are identified by numbers, the field of view is 8 arcsec, North is up, east to the left. In the right panel selected spectra extracted from the 3D data. At the top combined and normalized spectrum of the clusters 3, 4, and 8. The extraction regions of the other spectra are indicated by arrows.}
\label{3125_K}
\end{figure*}

The continuum emission of the field observed in NGC~3125 in the H and K band is dominated by one bright object corresponding to the center of the star forming region A. This source is significantly elongated in the EW direction and has a FWHM of 2.5$\times$4 pixels (0.3$\times$0.5 arcsec). In addition to this bright source we detect a number of fainter nearby compact sources: in Fig. \ref{3125_K} we show the K continuum image with the sources detected identified by numbers according to decreasing K brightness.
 We extracted the photometry of source A from the spectra using the transmission curves of the
 H and Ks filters and an aperture of radius 6 pixel, equivalent to 0.75 arcsec, and subtracted the background measured on a surrounding anulus. The measured magnitudes are H=15.6 and K=15.0.
For the same source Vanzi \& Sauvage (2006) found K=14.81, while Vanzi, Hunt \& Thuan (2002)
 K=14.97, H-K=0.47. 
The agreement is good considering that the previous observations were performed with a lower angular resolution. The fainter sources have K magnitudes around 18-19 and a
color H-K around 0.5 - 0.8. The aperture photometry of these sources is reported in Tab. \ref{3125_phot}. All sources observed appear resolved and have similar FWHM of about 3 pixels or $\sim$ 0.4 arcsec. Unfortunately we do not have a precise determination of the complex PSF produced by the AO system and we cannot deconvolve for it, therefore the values measured can only be considered upper limits, and
they are equivalent to about 170 pc diameters. This size is on the upper side for super massive stellar clusters, so we can tentatively refer to these objects as super stellar clusters. The sizes measured in the HST images for A1 and A2 are much smaller, 0.17 and 0.13 " respectively. The larger sizes measured in the infrared may be due to an envelope of warm dust around the stellar cluster. 
Clusters A1/2, 5, 6 and 7 show a progression of ages from few to about 10 Myr as indicated by the equivalent width of Br$\gamma$. 
We do not detect any emission lines in cluster 3, 4, and 8, so they are the oldest clusters observed in the field. They are also farther from source A than clusters 6 and 5.

\begin{table}
\begin{center}
\begin{tabular}{llll}
\hline
source  &   H    &   K     & Br$\gamma$ eqw  \\
\hline
 A (1-2)     &  15.6   &  15.03    &  210    \\
                  &             &   14.81$^*$  &       \\
                  &             &   14.97$^{**}$  &   \\
      3          &  18.9  &    18.1    &  -    \\
      4         &  19.4   &    18.5    &  -    \\
      5         &  19.4   &    18.7    &  160  \\
      6         &  19.2   &     18.6   &  130  \\
      7         &   19.4  &     18.8   &    50  \\
      8         &   20.0  &     19.4   &    -     \\
\hline
\end{tabular}
\caption{Aperture photometry and equivalent width of Br$\gamma$ for the clusters in NGC~3125. The aperture photometry was extracted with an aperture of 0.5 arcsec diameter, with the exception of cluster A (1-2) for which an aperture with diameter of 1.5 arcsec was used. $^*$ value from Vanzi \& Sauvage (2006), $^{**}$ value from Vanzi et al. (2002). Equivalent width in $\AA$.}
\label{3125_phot}
\end{center}
\end{table}

\begin{figure*}
\centering
\includegraphics[angle=0, width=16cm]{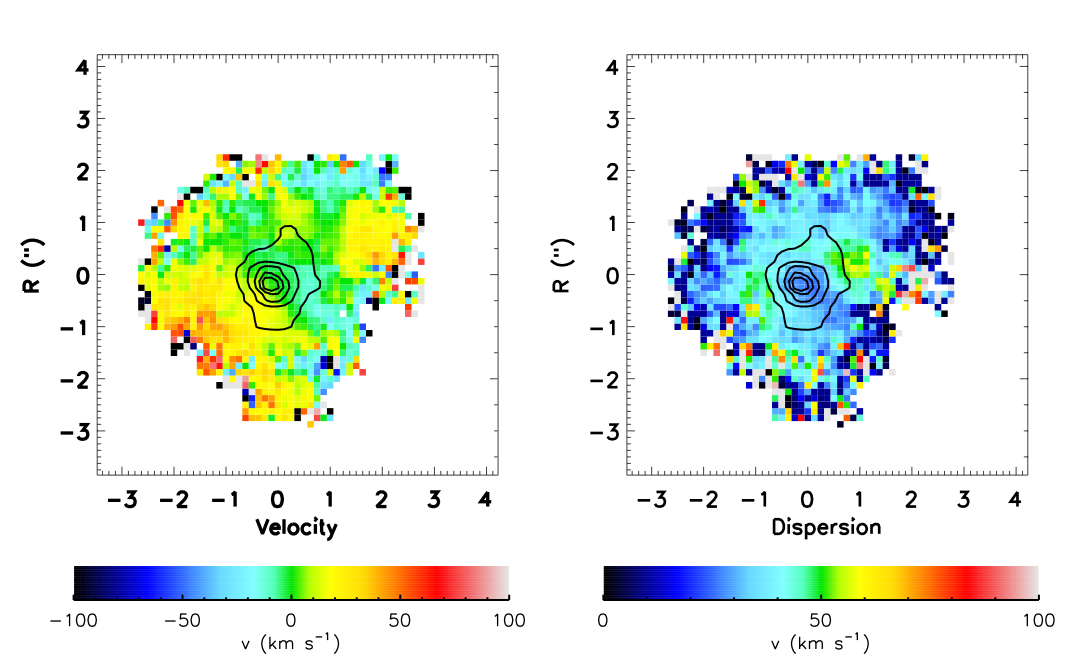}
\caption{Br$\gamma$ velocity field (left panel) and velocity dispersion (right panel) in NGC~3125.The contours represent the Br$\gamma$ emission.}
\label{3125_vel}
\end{figure*}

We combined the spectra of these three clusters, in an attempt to increase the S/N, and have a weak detection of the stellar CO features beyond 2.29 $\mu$m, compatible with the assumption that those are relatively evolved stellar clusters, older than 10 Myr.
The spectrum is shown in Fig. \ref{3125_K}.

\subsection{Ionized hydrogen}
The image obtained in Br$\gamma$ show the distribution of the ionized gas and identifies the main regions of star formation. We find that most star formation is concentrated in source A. 
The integrated Br$\gamma$ flux measured on this source with a radius of 0.5 arcsec is $4.5~10^{-15} erg/s/cm^2$ with an eqw = 210 \AA.
Comparing the Br$\gamma$ equivalent width with the results of SB99 we derive an age younger than 4 Myr. 
We extracted from our data a spectrum centered on cluster A with the same aperture as in Vanzi et al. (2002).
We then compared our spectrum with theirs and found good agreement between the fluxes measured in the two observations.

We marginally resolve source A 
into two components that we identify with A1 and A2 according to the HST observations presented by Hadfield \& Crowther (2006).

The bright Br$\gamma$ source is shifted by about 0.3 arcsec toward the 
east with respect to the peak of the brightest source of K continuum. 
This is easily explained with a picture of source A consisting of two clusters, A1 and A2 separated by 0.5 arcsec as given by HST images.
While these sources are of comparable brightness in the K continuum, A1 dominates the line emission and it is much brighter in the UV  (Hadfield \& Crowther 2006). 
We therefore identify the peak of the Br$\gamma$ flux with A1. The peak of the K continuum falls approximately between A1 and A2.

A region of diffuse and much fainter Br$\gamma$ emission is detected around source A extending with plumes across our full field of view. Our observations, however, do not reveal the presence of other prominent compact sources.

The total Br$\gamma$ luminosity observed over our field of view, obtained by integrating on a circular aperture of radius 23 pix (2.87"), is $1.97~10^{38} erg/sec$. Correction for extinction (Section 3.3) produces a value of $3.47~10^{38} erg/sec$.
This luminosity requires about 3300 equivalent O7V stars in terms of ionizing flux. Hadfield \& Crowther (2006) estimate 2000 of such stars based on the H$\alpha$ flux. This difference indicates how in the near IR we observe deeper 
into
 the star forming region. We used SB99 (Leitherer et al. 1999) to calculate the total number of O stars required and obtained  about 15500.


Assuming a temperature of 10,000 K for the ionized gas and Case B recombination theory we
derive a volume emission measure $(N_e)^2V$ of $8.0 \times 10^{64}$ cm$^{-3}$ from the 
measured Br$\gamma$ emission. This is a factor of 270 less than the volume emission
measure found by Thompson et al. (2009) for the sum of all of the sources in SBS0335-052E.
This indicates that the star formation rate in SBS0335-052E is significantly greater
than in NGC 3125.




In Fig. \ref{3125_vel} (left panel) we present the velocity field derived from Br$\gamma$, the Br$\gamma$ contours are overplotted on the image. The center of source A is set at 0 km/s and the velocity field is measured with respect to this reference. We observe possible evidence for a weak rotation around an axis crossing the galaxy in the NE to SW direction. Irregular patches with velocities up to +50 and down to -50 km/s are detected through the galaxy. These values are typical of the winds of ionized gas observed in blue dwarf galaxies (Martin 1999), though the interpretation of our observation in this sense is not straightforward based on the data available.
The line width of Br$\gamma$ is typically of few tens km/s, see the velocity dispersion map on the right panel of Fig. \ref{3125_vel}. The spectral resolution is not high enough to detect any particular structure in velocity dispersion or in the line profiles that might provide a detailed study of the velocity structure, only a slightly higher dispersion is seen in an area to the NW of the center.

\subsection{Dust and Extinction}

From the ratios of H$\alpha$, H$\beta$, and H$\gamma$, Hadfield \& Crowther (2006) measured E(B-V) = 0.24 and 0.21 respectively on 3125-A and 3125-B, these values are equivalent to $A_V$=0.74 and 0.65. Vacca \& Conti (1992) measured $A_V$=1.24 and 1.98 from H$\alpha$, H$\beta$. Galactic reddening is 0.25 mag.
Vanzi et al. (2002) measured the reddening from the ratio Pa$\beta$ / Br$\gamma$ and obtained $A_V$=1.6 and 3.3 respectively on 3125-A and 3125-B. 

Here we use the Br$\gamma$ and Br12 images to measure the extinction. We assumed an intrinsic ratio Br12/Br$\gamma$=0.19 corresponding to case B, $T_e=10^4 K$, and $n_e=10^2 cm^{-3}$ (Hummer \& Storey 1987). We extracted spectra centered on A with a circular aperture of radius 8 pixel and measured $A_V$ = 4.8 using the extinction law of Rieke \& Lebofsky (1985) and $A_V$=5.4 using the extinction law of Calzetti et al. (2000). We assumed the latter value considering the Calzetti's law more suitable for our case.

We obtained an extinction map but it does not show any particular structure except for a trend of higher extinction in correspondence
of source A. The S/N is too low to determine any extinction structure beyond this.
 

We have already
 proved that our relative H to K calibration is reasonably good and in agreement with data from the literature, so we are confident that the results derived from the comparison of H and K spectra are reliable.

\subsection{Molecular Hydrogen}
\begin{figure}
\centering
\includegraphics[angle=0, width=8cm]{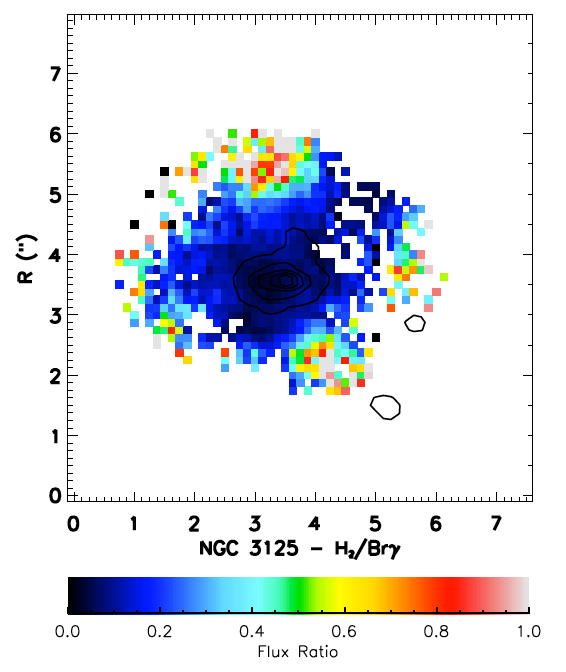}
\caption{Ratio of the line $H_2$ at 2.12 $\mu m$ and Br$\gamma$ in NGC~3125. }
\label{3125_H2}
\end{figure}

The image obtained in the line of $H_2$ (1-0)S(1) at 2.12 $\mu m$, rest wavelength, gives a picture of the distribution of the molecular hydrogen in NGC 3125-A. This is very different from the morphology observed in Br$\gamma$ and in the continuum. The flux is relatively faint at the position of source A, while the diffuse emission extends further to the north and to the south. To facilitate the comparison in Fig. \ref{3125_H2} we display the ratio $H_2$/Br$\gamma$. The ratio is lower toward the center of region A and higher further out indicating how the molecular hydrogen emitting region reaches beyond the ionized atomic hydrogen. We also see in Fig. \ref{flux_3125} how the $H_2$ emission forms a bow like structure extended in the N-S direction and with its convexity toward the E.

A large number of hydrogen molecular lines are detected in adition to (1-0)S(1). In Tab. \ref{3125_h2} we report the measurements obtained 
from
 a spectrum centered on the region A. The comparison of the ratios with models (Black \& van Dishoeck 1987) indicates fluorescence as the main mechanism of excitation. We took the line (2,1)S(1) 
as a 
reference because it is the brightest of those sensitive to the excitation mechanism and made a map of the ratio (2-1)S(1)/(1-0)S(1).
We found the ratio to be roughly constant at around 0.5 over the entire area observed, consistent with fluorescence.
We infer therefore that UV photons from young massive stars are the main source of excitation of molecular hydrogen through out the area observed.
The $H_2$ emitting region is shaped by the penetration of the UV photons and by the extent of the molecular cloud.

\begin{table}
\begin{center}
\begin{tabular}{llll}
\hline
line  &   line/(1-0)S(1)    &   line   & line/(1-0)S(1)  \\
\hline
 (1-0)S(3)  &  0.30     &   (2-1)S(1)   &    0.5   \\
 (1-0)S(2)  &  0.25     &   (2-1)S(0)   &    0.5   \\
 (2-1)S(3)  &  0.32     &   (1-0)Q(1)   &    1.7   \\
 (2-1)S(2)  &  0.17      &  (1-0)Q(2)   &    0.45 \\
 (1-0)S(0)  &  0.45     &    (1-0)Q(3)  &    0.8    \\
\hline
\end{tabular}
\caption{Molecular hydrogen lines detected in NGC~3125 and ratios to (1-0)S(1). The errors on the ratios are smaller than about 20\%.}
\label{3125_h2}
\end{center}
\end{table}

\begin{figure}
\centering
\includegraphics[angle=0, width=8cm]{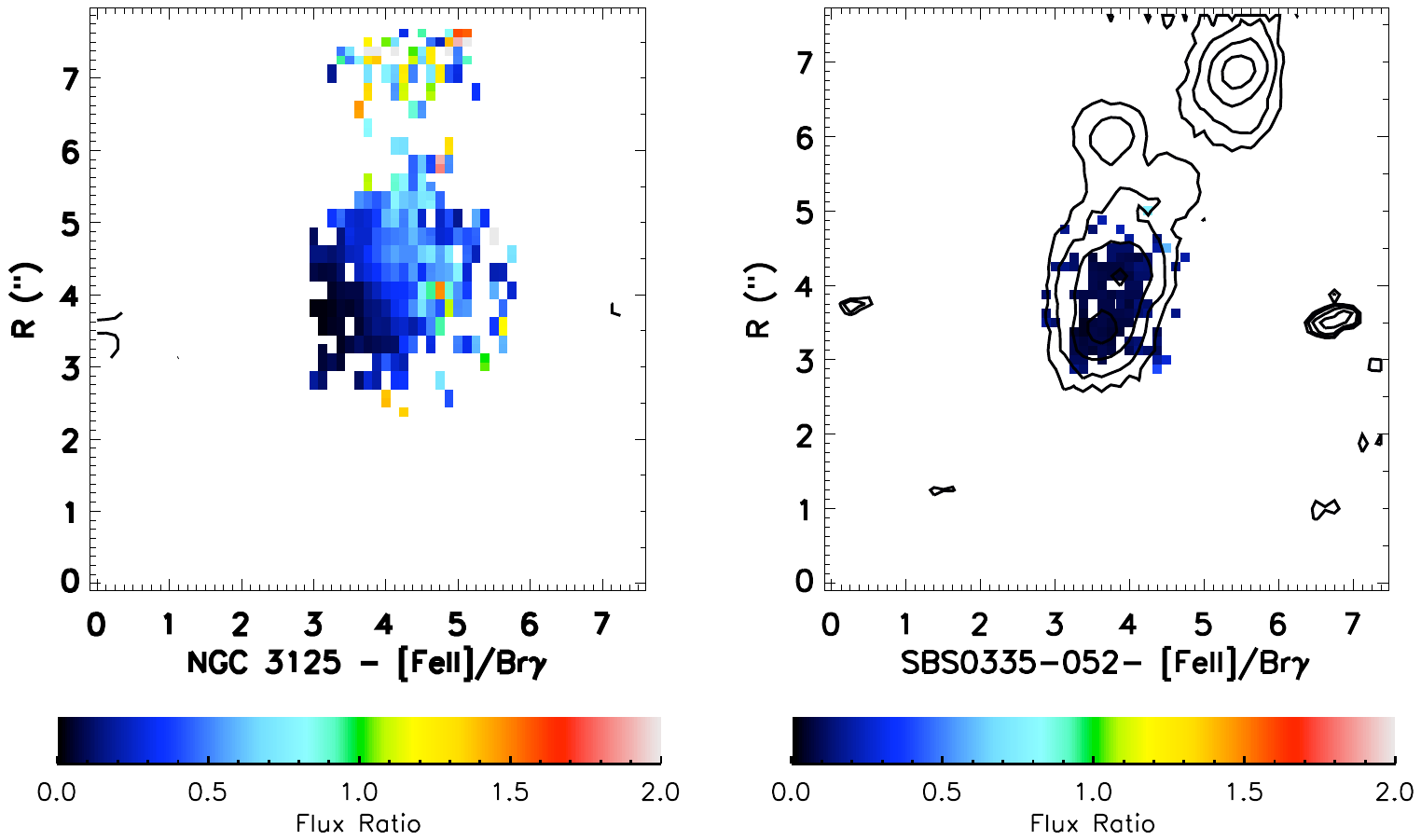}
\caption{Map of the [FeII]/Br$\gamma$ ratio for NGC~3125}
\label{FeII_3125}
\end{figure}

\begin{figure}
\centering
\includegraphics[angle=0, width=6.5cm]{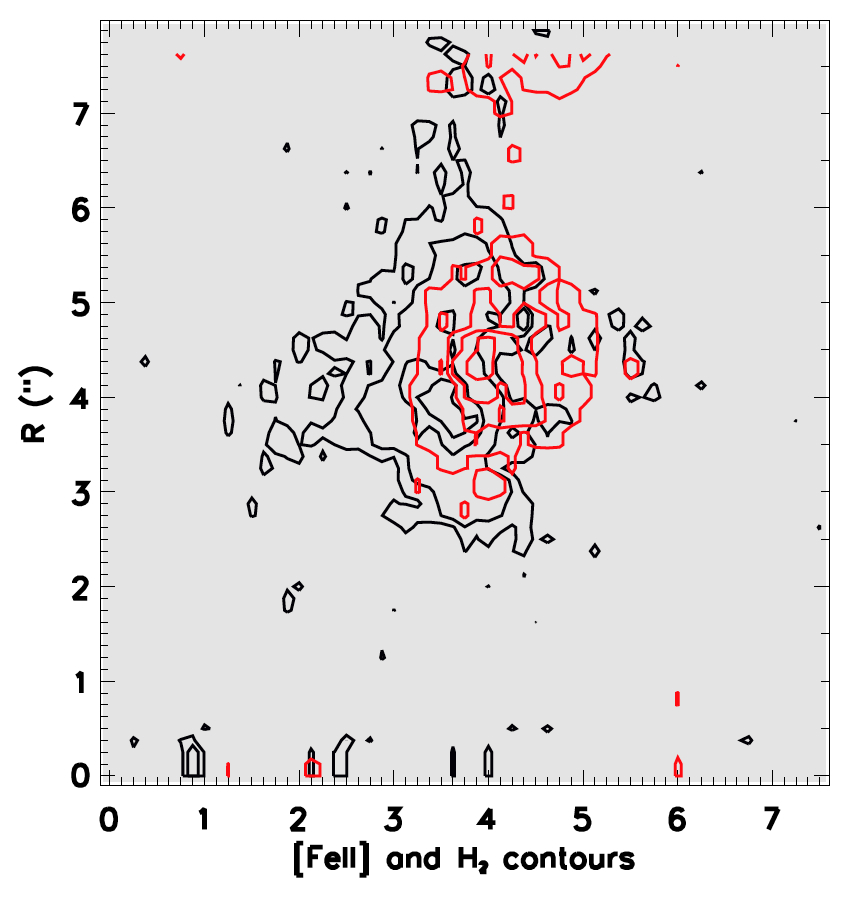}
\caption{Contours of $H_2$ (black lines) and [FeII] (red lines) in NGC~3125.}
\label{3125feiih2}
\end{figure}

\subsection{Ionized Iron}

More surprises are to be found when observing NGC~3125-A in the line of forbidden iron at 1.64 $\mu m$. At this wavelength in fact the emission of the galaxy is shifted toward the north west by about 1 arcsec with respect to the underlying continuum. To make the difference more visible in Fig. \ref{FeII_3125} we show an image of the [FeII]/Br$\gamma$ ratio. The absolute value of the ratio is obtained re-scaling [FeII]/Br12 to  Br$\gamma$, in this way we remove any systematic error produced by the H/K relative calibration and minimize the effect of extinction. The values span a range from about 0.06, which is typical of photoionization, to values higher than 2 indicating significant contribution from SN remnants.

The area of bright [FeII] emission is almost 2" long in the NS direction and more than 1" wide in the EW direction.
It is close (0.4 arcsec) to the area of high velocity dispersion observed in Br$\gamma$ (see Fig. \ref{3125_vel}).
According to Alonso-Herrero et al. (2003) it is possible to use the [FeII] flux to estimate the number of SN remnants present in the observed field by comparing the [FeII] luminosity with the estimated luminosity of a SN remnant.
The [FeII] flux observed is $9.14~10^{-16} erg/s/cm^2$, which gives a luminosity of $3.2~10^{37} erg/s$ corrected for extinction. From this value, using the calibration of $8~10^{36} erg/s/SN$ of Vanzi \& Rieke (1997) we estimate about 4 SN remnants. Assuming a [FeII] bright phase of about $10^4$ yr we obtain a SN rate of $4~10^{-4}~yr^{-1}$. 
The relation (1) of Cresci et al. (2010), which relates the [FeII] luminosity with the SN rate through a constant  0.24 $(10^{40} erg/s yr)^{-1}$, gives instead $7~10^{-4}~yr^{-1}$.
These values can be compared with the SN rate derived by Stevens et al. (2002) for NGC~3125-A from radio 
observations of
$4~10^{-3}~yr^{-1}$, exactly 10 times higher. 
The sub-solar metallicity of the galaxy is not enough to explain the difference, as it is only a  factor of about 3.
Besides observational and calibration errors there is the possibility that even in the IR we do not see all the SNe detected in the radio.
This may be due to extinction as proposed by Neff et al. (2004), Cresci et al. (2007), and P{\'e}rez-Torres et al. (2009).

It is interesting to compare the morphology of the [FeII] with $H_2$. In Fig. \ref{3125feiih2} we show the contours of the two lines. We observe that the [FeII] emission almost 
fills the cavity left by the bow like structure of $H_2$ described previously.
This observation will be discussed further in Section 5.

We detect two lines of [FeIII] at 2.217 and 2.347 $\mu$m but at a much lower signal-to-noise, preventing us from obtaining reliable emission maps.

\section{Analysis of SBS~0335-052}

\begin{figure*}
\centering
\includegraphics[angle=0, width=19cm]{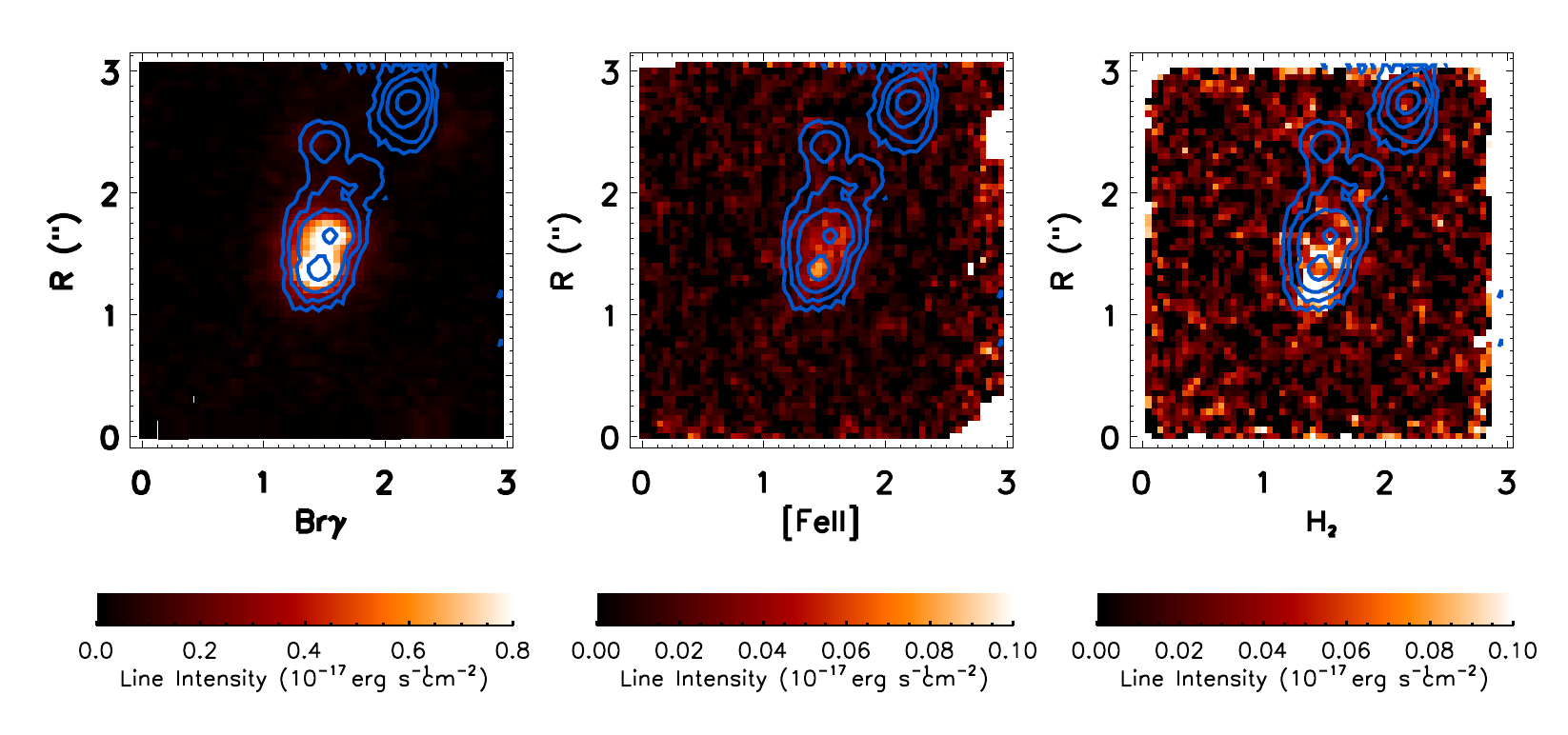}
\caption{Br$\gamma$, [FeII] and $H_2$ images of SBS~0335-052, the contours of the continuum are over plotted on each image. The continuum plotted corresponds to band H for [FeII] and to band K for Br$\gamma$ and $H_2$.}
\label{flux_sbs}
\end{figure*}

In Fig. \ref{flux_sbs}  we show the images in Br$\gamma$ 2.167 $\mu$m, $H_2$ 2.121 $\mu$m, and [FeII] 1.643 $\mu$m rest wavelengths of SBS~0335-05. The contours of the continuum are overplotted for reference. We will use these images as the base of our analysis as previously done.

\subsection{Near Infrared continuum}
\begin{figure}
\centering
\includegraphics[angle=0, width=8cm]{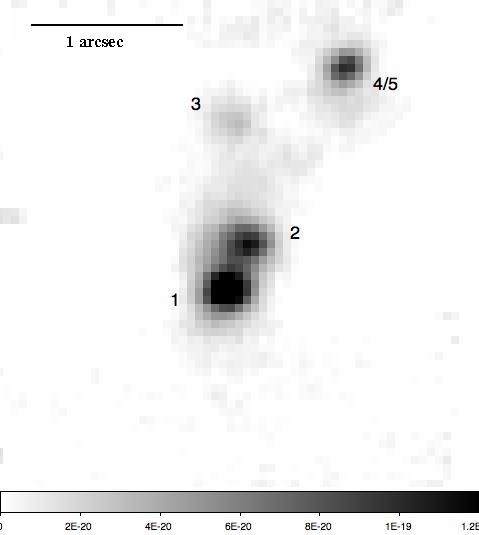}
\caption{SBS~0335-052 in the K continuum, the compact sources detected are identified by numbers, the field of view is 3.2 arcsec, North is up, east to the left.}
\label{sbs_K}
\end{figure}

A number of compact sources are detected in the H and K continuum mainly those named S1, S2, S3, and S4 by Thuan et al. (1997), see Fig. \ref{sbs_K}. We do not detect S7 (
Thompson et al. 2009), while S6 (Thuan et al. 1997) and S8 (Thompson et al. 2009) fall out of our field of view. We compared the aperture photometry obtained on the sources detected in our observations with data from the literature similarly to Section 3.1 and found our observations fully 
consistent with previous ones. The results are reported in Tab. \ref{sbs_phot}. 

\begin{table}
\begin{center}
\begin{tabular}{lllll}
\hline
source  &   H (0.40)   &   H (0.88)   & K (0.40)  & K (0.88)   \\
\hline
1-2     &  19.52   &  18.56  &  18.24  (18.45)  &  17.41  (17.55) \\
4-5     &  19.78   &  19.19  &   19.31  (19.32)  &  18.84  (18.37) \\
\hline
\end{tabular}
\caption{Aperture photometry of clusters in SBS~0335-052. The apertures are indicated as diameters in arcsec. in parenthesis we report the values from Vanzi et al.( 2000)}
\label{sbs_phot}
\end{center}
\end{table}

These sources span a sequence age of about 25 Myr according to Thuan et al. (1997) or 15-20 Myr according to Thompson et al. (2009) running from the north to the south with the oldest objects in the north near a SN bubble that suposely triggered the star formation observed today.
Since S4 is not detected in Pa$\alpha$ or Br$\gamma$ it is certainly older than 10 Myr and, in principle, should show features of a relatively evolved object.
The H and K spectra of this object do not present any evidence of stellar absorption bands.
In particular the CO bands at 2.29 $\mu$m are completely absent.
We attribute this to the lack of metals in the galaxy. 
To quantify this better we use the results of SB99 (Leitherer et al. 1999) even though the model only reaches down to a metallicity of about 1/20 of solar.
For this metallicity the CO index is basically 0 at an age of 10 Myr and less than a factor 2 below the solar value at an age of 30 Myr.

\subsection{Ionized Hydrogen}
\begin{figure}
\centering
\includegraphics[angle=0, width=8cm]{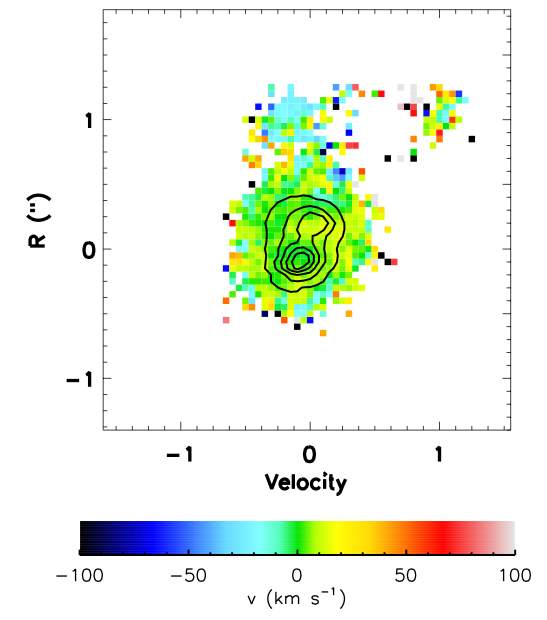}
\caption{Br$\gamma$ velocity field in SBS~0335-052 with Br$\gamma$}  contours overplotted.
\label{sbs_vel}
\end{figure}

In Br$\gamma$ we detect the same sources detected by Thompson et al. (2009) in Pa$\alpha$ mainly S1, S2, S3-Pa shifted to the 
east of S3, and S7 to the west of S4 as it can be seen in the left panel of Fig. \ref{flux_sbs}. To their analysis we can add  the velocity field of the gas, showed in Fig. \ref{sbs_vel}. The map is refered to 0 velocity on S1. Cluster S2 has a receding velocity of about 20-30 km/s, while S3 has a similar velocity in the opposite direction. A part from this relative motion of the clusters no other velocity structure can be detected.

\subsection{Dust and Extinction}
The presence of dust in SBS~0335-052 is certainly a 
complicated
 matter whose study gained relevance with the finding of Thuan et al. (1999) of an unexpected mid-IR excess attributed to thermal emission by dust and considered responsible 
for
a $A_V$ extinction of about 20 mag. The dust emission was later associated with source S1-S2 and the extinction measured directly $A_V$=12 mag (Dale et al. 2001, Hunt et al. 2001). Hirashita et al. (2002) considered the presence of dust in low metallicity environment 
to be consistent
with the amount of dust generated by type II SNe. The interpretation of the SED is far from being settled (Houck et al. 2004). 
Reines et al. (2008) and Johnson et al. (2009), based on the comparison of extinction and ionizing fluxes measured at different wavelengths, provide convincing evidence for a porous and clumpy ISM surrounding the natal clusters.

Previously Izotov et al. (1997) measured $A_V$=0.55 from the Balmer decrement toward S1-S2.
Vanzi et al. (2000) measured $A_V$=0.78 from H$\beta$/Br$\gamma$ and $A_V$=0.55 from H$\alpha$/H$\beta$.
Thompson et al. (2006) measured $A_V$=12 from Br$\gamma$ and Br$\alpha$.

Our data are good enough to measure the extinction from the Br12/Br$\gamma$ ratio towards S1, S2 and S3.
For this purpose we extract spectra centered on each of these sources with radius of 3 pix.
For all clusters we find consistently the same value: $A_V$=3.0 mag. with the extinction law of Rieke \& Lebowsky (1985) and $A_V$=3.4 using the law of Calzetti et al. (2000).

We generated an extinction map over the entire Br$\gamma$ emitting region and found uniform extinction across it. 
This seems to indicate an extended absorbing cloud in which the star forming centers are embedded.
We can then speculate that the clumpiness of the ISM must be on a scale smaller than our angular resolution.

 For the extinction correction purpose we assume the picture of Reines et al. (2008) and correct for the value in magnitude measured at each wavelength + 1 mag. for wavelengths shorter than Br$\alpha$, to take into account the 60\% covering factor product of dense dust clumps.
 
\subsection{Molecular hydrogen}
Molecular hydrogen is detected in SBS~0335-052 only at the position of S1 and S2. Our new observations are mostly consistent with those of Vanzi et al. (2000) in all respects. The line ratios are all consistent with theirs and consistent with florescent excitation. Differently from them we do not detect the line (2,1)S(0). We compare carefully the spatial distribution of $H_2$ with Br$\gamma$ and [FeII] and differently from NGC~3125 we cannot detect any significant deviation. We measured the $H_2$ 2.121 flux in S1 and S2 with the same apertures as Thompson et al. (2009) - 0.38" diameter - but obtained a total flux smaller by a factor 2. We attribute this difference to the combination of lower Strehl ratio provided by our AO observations, compared to HST, and the small apertures.

\subsection{Ionized Iron}
We detect the [FeII] 1.64 forbidden emission line at both S1 and S2.
Away from these two sources the flux drops and we do not detect the line at S3 or at the position of any of the other clusters.
The forbidden line of [FeII] 1.64 was not detected in SBS~0335-052 in earlier observations (Vanzi et al. 2000) and its absence interpreted 
to be consistent
 with the young age of the star forming episode. Our detection of the line, mostly due to the enhanced spectral resolution, allows some further consideration. 

The ratio [FeII]/Br$\gamma$ on S1 and S2 is respectively 0.065 and 0.085 which was measured, as previously discussed, by rescaling from Br12. The error on the ratio is about 0.01.
These values are consistent with photoionization by the clusters S1 and S2 at solar metallicity, however in our case they certainly 
exceed what should be expected because of the metal poor environment. For instance the models and measurements presented by Lanfranchi \& Matteucci (2010) would suggest that the iron 
has a relative abundance similar to oxigen. 
From the total flux observed by us in an aperture of 1.4" in diameter, which includes both S1 and S2, of $7.4~10^{-17} erg/s/cm^2$, we find a luminosity of $8.6~10^{37}$ erg/s, corrected for an extinction of $A_{1.64}$=0.36 + 1 mag. 
To estimate the photoionized contribution to [FeII] we can multiply the Br$\gamma$ luminosity by a factor 0.06 which is the [FeII]/Br$\gamma$ photoionized ratio in 
a solar
 metallicity environment. 
The Br$\gamma$ observed flux of S1-S2 is $1.24~10^{-15}$ erg/s/cm$^2$ which is equivalent to a luminosity of $1.2~10^{39}$ erg/s when corrected for an extinction of $A_{2.16}$=0.18 + 1 mag.
It, therefore, gives a photoionized [FeII] luminosity of $7.2~10^{37}$ erg/s which still must be scaled down to the low metal abundance of the galaxy.
For this purpose we assume that iron scales as oxygen (Lanfranchi \& Matteucci 2010) and apply a factor 1/40, giving $1.8~10^{36}$ erg/s. 
If this is correct photoionization provides only a small fraction (2\%) of the observed [FeII]. Hunt et al. (2004) detect a radio continuum with a non thermal component which they associate 
with
 the clusters S1 - S2 and consider consistent with the previously determined SN rate of $6~10^{-3} SN/yr$ (Hunt et al. 2001). However Johnson et al. (2009) find only marginal evidence for the presence of a non-thermal radio component. We wish
to establish whether our observation is consistent with the SN rate reported in the literature.
To do so we refer to relation (1) of Cresci et al. (2010), which relates the [FeII] luminosity to the SN rate using, as they do, a constant $\alpha = 0.24 (10^{40} erg/s~yr)^{-1}$.

The [FeII] luminosity of $8.6~10^{37}$ erg/s would provide a SN rate of $2~10^{-3}$ SN/yr. The metallicity issue is even more complex in this case as, beside the abundance of the ISM, there is the unknown effect of the SN ejecta. In any case the SN rate calculated must be considered a lower 
limit,
possibly to be scaled up by a factor 40. If we do so we end with 0.08 SN/yr, a factor 13 higher than the prediction of Hunt et al. (2001). These considerations would point to a bulk of the [FeII] emitted by SN remnants, with a possible excess of [FeII] that could be explained through the poorly known effect of metallicity or with a possible contribution of shocks generated by high stellar winds (Thompson et al. 2006).

In this picture the complete lack of [FeII] at the position of clusters S3, S4 and S5 is surprising.
We can try to estimate the expected [FeII] flux from these 
sources by re-scaling the measurements of S1-S2. 
According to the K continuum measurements obtained by us and by Thompson et al. (2009) we can estimate that S3, S4, and S5 are less massive than S1 and S2 by a factor 10 or less.
It is almost certainly less if we consider that the S1-S2 continuum must receive a nebular contribution from the ionized gas, while S3, S4, and S5 do not.
Based on the K continuum we can assume that the S4-S5 
complex is about 3 times less massive than S1-S2. 
We can also make a conservative assumption that half of the [FeII] luminosity from S1-S2 is photoionized even though a fraction of less than few \% is certainly more realistic.
We are then left with $4.3~10^{37}$ erg/s divided by a factor 3 to take into account the mass difference.
The SN remnants, by conservative rescaling of the S1-S2 observation, then place lower limit on the S4-S5 [FeII] emission of about $1.4~10^{37}$ erg/s.

Our line detection limit in a circular aperture of radius 3 pix is about $10^{-18} erg/s/cm^2$, equivalent to a line luminosity of $3.3~10^{35} erg/s$. This is about 40 times lower than the expected lower limit on the luminosity of S4-S5.
Why don't we observe [FeII] in the evolved clusters? The low abundance cannot be advocated in this case because we started our reasoning from values observed in the very same galaxy.
We have two possible explanations: 
i) our mass re-scaling factor for S4-S5 is largely overestimated
ii) S4-S4 are older than 40 Myr, or both. 
Our non detection of [FeII] at the location of evolved clusters goes together with the weak evidence for non-thermal radio emission.
The situation of S4-S5 seems similar to clusters 3, 4, and 8 in NGC~3125.
Those clusters, however, are more than 3 magnitudes fainter in K than source A.


\section{Discussion}
In NGC~3125 we detect different sources of near infrared radiation. The continuum is mostly concentrated in one bright source, surrounded by a number of significantly fainter objects. We may consider these as a population of 
clusters that are relatively evolved. 
The brightest source is a large region of recent star formation where there is more than one cluster present, two of which that we associate with the HST clusters A1 and A2.

The ionized gas is concentrated in one single source, HST cluster A1, which can be considered the main ionizing source of the entire region.
 This must be a young and massive cluster.

The ionized iron peaks at a position different from clusters A1 and A2 located about 1 arcsec to the north-west with an observed flux that is consistent with 4 SN remnants.
This area must be the location of a previous episode of star formation with an age of less than 40 Myr. It is likely that the older star formation triggered the most recent one. Successive episodes of star formation triggering one another have been 
hypothesized
 (Cresci et al. 2010). This idea finds further support in the comparison of the distribution of [FeII] relative to $H_2$. We believe that the molecular hydrogen is mostly photoexcited by the young clusters and its distribution delineates somehow the molecular cloud which is currently undergoing star formation. We also 
show how this cloud forms a bow like structure with a cavity and how the emission of [FeII] almost fills that cavity, Fig. 6. If the [FeII] traces the effect of SN remnants produced in previous star formation we have a direct evidence of ISM compression and SF triggering by 
SNe. The distance between the center of [FeII] emission and the current star formation is about 1" or 55 pc, equivalent to a traveling time of about 1 Myr for a shock speed of 50 km/s. The time required to produce the first 
SN is aproximately 6 Myr, giving in total a delay time of about 7 Myr.

We may tentatively
 identify in NGC~3125 three episodes of massive star formation closely separated in time: the youngest episode that produced sources A1/2, 5, 6 and 7, a previous episode at the location of the bright [FeII] source, and the older clusters A3, 4 and 8.

In SBS~0335-052 we identify the population of clusters detected by HST in the visible and near-IR. We detect molecular hydrogen 
that is mostly fluorescently excited. The picture provided is fully consistent with previous observations. The study of the dynamics provides the picture of a relatively quiet
 galaxy. We detect for the first time the [FeII] forbidden line and we have difficulties in interpreting it quantitatively mainly because of the relatively poorly known effects of the 
metal abundance and of the [FeII] luminosity of SN. Based on what we believe reasonable hypothesis we find lack of [FeII] at the location of the most eveolved clusters, possibly indicating for them an age older than previously though.

 If we compare all galaxies observed so far mainly, II~Zw~40 (Vanzi et al. 2008), NGC~5253, He~2-10 (Cresci et al. 2010), NGC~3125 and SBS~0335-052, we can highlight the following points:
 
 1) In all galaxies we observe young massive clusters. A few of them are younger than 2-3 Myr and are surrounded by large HII regions. Several others show evidence of star formation in the past few tens Myr;
  
 2) The youngest clusters are also surrounded by areas of warm $H_2$ emitting in the near IR. These areas are much more extended than the HII regions in some cases such as NGC~3125, II~Zw~40 and NGC~5253. In one case (II Zw 40) we could associate the $H_2$ to the CO molecular cloud observed in the mm. The $H_2$ is predominantly fluorescently excited. The case of SBS~0335-052 is somehow different as, at the angular resolution reached so far, the $H_2$ seems exactly associated to the HII region;

3) the regions of [FeII] emission appear spatially associated with the HII regions when photoionized as in II Zw 40. However, when we observe evidence for shock excited [FeII] this appears to be spatially far from the HII region as in He~2-10 and NGC~3125. This picture is also observed in NGC 1569 (Pasquali et al. 2011). In SBS~0335-052 the [FeII] is presumably shocked and at the same location of the HII regions, however our spatial resolution is significantly lower for this galaxy. We can use the shock excited [FeII] as tracer of recent star formation, because SN would be present up to about 40 Myr after the start of an instantaneous burst of star formation. We therefore conclude that the formation of massive clusters, once triggered, migrates as SN appear, propagating itself through the galaxy where fresh molecular gas is available.

\section{Conclusions}
We obtained integral field spectroscopy in the near infrared of two blue dwarf  galaxies: NGC~3125 and SBS~0335-052. The study of the spectral features observed provided a detailed view of the environment where the star formation is occurring. We found that:

   \begin{enumerate}
      \item NGC~3125 hosts a number of massive clusters spanning a range of ages, a similar result was already known for SBS~0335-052.
      
      \item The molecular hydrogen is mostly fluorescently excited in both galaxies, however while in NGC~3125 it extends beyond the giant HII region around the youngest clusters, in SBS~0335-052 it appears to be
 concentrated in the giant HII regions associated to young clusters. The difference however could be only apparent due to the larger distance of SBS.
      
      \item We detect [FeII] emission in both galaxies. In both cases the [FeII] luminosity is roughly consistent with the evolutionary stage of the galaxies and the SN rate derived from radio observations, once the effect of abundance is taken into account. In NGC~3125 the [FeII] emitting region is significantly separated from the HII region and the molecular hydrogen emitting region while in SBS~0335-052 it mostly coincides with them. Again this coincidence could be only apparent and produced by the larger distance of SBS.
In SBS~0335-052 we do not detect [FeII] from relatively evolved clusters where we would have expected it.
      
      \item The dust distribution is mostly uniform in both galaxies with 
a slightly higher concentration at the 
positions of the youngest clusters.
      
      \item When considering all galaxies observed by us and similar results available in the literature in all of them we find evidence for
self propagating star formation whose time scale is consistent with the delay required to form 
SN at each location and a shock 
propagation speed of few tens km/s.

    \end{enumerate}

\begin{acknowledgements}
This research has made use of the NASA/IPAC Extragalactic Database (NED) which is operated by the
Jet Propulsion Laboratory, California Institute of Technology, under contract with the National Aeronautics and Space Administration. 
LV received support from CONICYT through project Fondecyt n. 1095187 and project Anillo ACT-86. GC acknowledges support by the ASI-INAF grant  I/009/10/0.
\end{acknowledgements}

\end{document}